\documentclass[prb,showpacs,twocolumn,preprintnumbers,amsmath,amssymb]{revtex4-1}
\usepackage{graphicx}

\begin{document}

\title{Scattering matrix of elliptically polarized waves}

\author{M. Mart\'{\i}nez-Mares}
\affiliation{Departamento de F\'{\i}sica, Universidad Aut\'onoma
Metropolitana-Iztapalapa, A. P. 55-534, 09340 M\'exico D. F., Mexico}

\author{E. Casta\~no}
\affiliation{Departamento de F\'{\i}sica, Universidad Aut\'onoma
Metropolitana-Iztapalapa, A. P. 55-534, 09340 M\'exico D. F., Mexico}

\begin{abstract}
We analyze the scattering of elliptically polarized plane waves normally incident at the planar interface between two different materials; we consider two cases: dielectric-dielectric and dielectric-conductor interfaces. The scattering matrix $S$ in both cases is obtained using the boundary conditions and Poynting's theorem. In the dielectric-dielectric case we write $S$ using two different basis, the usual $xy$ and a rotated one. For the dielectric-conductor interface, the use of the rotated basis together with an energy balance argument leads us, in a natural way, to construct a unitary $S$ matrix after recognizing the need to introduce two equivalent parasitic channels due to dissipation in the conductor, and the transmission coefficient into these parasitic channels measures the absorption strength.  
\end{abstract}

\pacs{41.20.Jb, 42.25.Bs, 42.25.Fx, 42.25.Gy, 42.25.Ja}

\maketitle


\section{Introduction}

The scattering of waves has attracted much attention even before being an object of scientific inquiry, since the observation of these phenomena is fascinating and even pleasurable. Take for example, the observation of water waves moving on the surface of a pond, where diffraction and interference effects give origin very interesting patterns. Nowadays, scattering phenomena are use in general to study how a wave an a target are transformed by their mutual interaction, in such a way that we extract very useful information about the structure of the target and the character of the wave itself.\cite{Crawford,Elmore}

In wave phenomena, either classical or quantic, the dispersion is mainly characterized, in an elegant and compact way, by the scattering $S$ matrix, that describes the transformation of an incoming wave into an outgoing one due to the interactions with a particular target.\cite{Newton,Mello} This has been used to discover the inner structure of many different objects ranging from macroscopic crystals, DNA molecules down to systems of atomic and nuclear sizes, and even smaller systems.~\cite{Kittel,Ashcroft}

In electromagnetic phenomena scattering is produced by variations either in time or space of the dielectric function;\cite{Jackson} if this function is real then $S$ is unitary since the energy flux is conserved. However, in the presence of a sink of energy the energy is not conserved as it happens in metals due to dissipation, in this case the dielectric function is complex and, therefore, $S$ is a sub-unitary matrix. A large majority of the work on scattering has been done with unitary $S$ matrices in problems where the flux is a conserved quantity, when dealing with problems where the flux is not conserved it then becomes natural to extend the $S$ matrix to a unitary one by increasing its dimensions to include dissipation channels.~\cite{Lewenkopf1992,Brouwer1997,vdom2008}

In a previous publication, \cite{vdom2008} we addressed the scattering of linearly polarized plane waves with normal incidence at the planar interface of two media for two different cases, one with a dielectric-dielectric interface, and another with a dielectric-conductor interface. This was done using Poynting's theorem at the interface to define an $S$ matrix; for the dissipative case, dielectric-conductor interface, the scattering matrix becomes sub-unitary, $\tilde{S}$, since the energy flux is not conserved. In this last case, we were able to write $\tilde{S}$ as part of a unitary $S$ matrix by the introduction of a single ``parasitic channel'' related to the energy dissipation in the conductor. However, that work only considered normal incidence of linearly polarized waves; in this work, we consider a more general case: elliptically polarized waves.  

In order to be self contained, in the next section we summarize the main ideas concerning Poynting's theorem. In Sect.~\ref{sec:dielectric} this theorem and the boundary conditions at the surface of two dielectrics to define a scattering matrix. Sect.~\ref{sec:conductor} is devoted to the dielectric-conductor interface where Poynting's theorem help us to extend the sub-unitary scattering matrix to a unitary one. 


\section{Energy balance equation}
\label{sec:Poynting}

Assuming a harmonic time dependence of the electric $\mathbf{E}(\mathbf{r},t)$, magnetic $\mathbf{H}(\mathbf{r},t)$, and density current $\mathbf{J}(\mathbf{r},t)$ complex fields, the time-averaged Poynting's theorem is given by the real part of the equation \cite{vdom2008}
\begin{equation}
\label{ec:cenergia2}
\nabla\cdot \mathbf{S}_P(\mathbf{r}) = -\frac 12
\mathbf{J}(\mathbf{r})\cdot \mathbf{E}^*(\mathbf{r}),
\end{equation}
where 
\begin{equation}
\label{eq:sprom2}
\mathbf{S}_P(\mathbf{r}) = 
\frac 12 \mathbf{E}(\mathbf{r})\times \mathbf{H}^*(\mathbf{r})
\end{equation}
is the time-averaged Poynting's vector. The right hand side of Eq.~(\ref{ec:cenergia2}) is  the negative of the time-averaged work done by the fields, per unit volume per unit time, and represents the conversion of electromagnetic energy to thermal (or mechanical) energy. The integral of Eq.~(\ref{ec:cenergia2}) over a volume $V$ enclosed by a surface $\Sigma$ gives (only the real part is physically relevant)
\begin{equation}
\label{eq:balance}
\Phi = -W, 
\end{equation} 
where, 
\begin{equation}
\label{eq:flux}
\Phi = \oint_{\Sigma}\mathbf{S}_P(\mathbf{r})\cdot\hat{\mathbf{n}}\,da ,
\end{equation} 
is the net flux $\Phi$ of $\mathbf{S}_P(\mathbf{r})$ through $\Sigma$, and 
\begin{equation}
\label{eq:W}
W = \frac 12 \int_V \mathbf{J}(\mathbf{r})\cdot \mathbf{E}^*(\mathbf{r})\, dV,
\end{equation} 
is the time-averaged rate of work done by the fields if there are dissipative processes in the system. 

\subsection{Energy flux conservation}

When there is no dissipation, $\mathbf{J}\cdot\mathbf{E}^*=0$, and therefore $W$ is zero, and the energy flux is conserved: 
\begin{equation}
\label{eq:conservation}
\Phi = 0. 
\end{equation}
This means that the net flux crossing into the system equals the one leaving it. 


\section{Dielectric-dielectric interface}
\label{sec:dielectric}

\subsection{Linear $xy$ polarization basis}

\begin{figure}
\includegraphics[width=0.7\columnwidth]{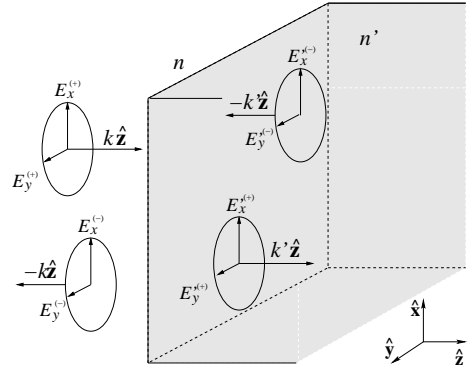} 
\caption{Elliptically polarized waves normally incident at the planar separation between two dielectrics with indices of refraction $n$ and $n'$, respectively.}
\label{fig:die-die1}
\end{figure}

Lets take the $xy$-plane as the surface that separates two dielectrics with indices of refraction $n$ and $n'$, as shown in Fig.~\ref{fig:die-die1}. We consider a normally incident plane wave which is the superposition of two linearly polarized waves, one in $\hat{\mathbf{x}}$-direction and the other one in $\hat{\mathbf{y}}$, what we call $x-y$ basis.

Therefore, the spatial part of the electric field for $z<0$ is given by
\begin{equation}
\label{eq:campos-n1}
{\bf E}(z) = 
\left( E_{ax} \hat{{\bf x}} + E_{ay} \hat{{\bf y}} \right) e^{ikz} + 
\left( E_{bx} \hat{{\bf x}} + E_{by} \hat{{\bf y}} \right) e^{-ikz} , 
\end{equation}
where $k=n\omega/c$; the subindex $a$ denotes incoming waves and $b$ outgoing ones. Since we are using an elliptically polarized wave we have that   
\begin{equation}
\label{eq:alpha-beta}
E_{ax} = \alpha\, E_a 
\quad \mbox{and} \quad 
E_{ay} = \beta\, E_a\, e^{i\phi_a},
\end{equation}
where $\phi_a$ is the phase difference between $x$ and $y$ components, being $\alpha$ and $\beta$ real numbers whose squares add up to one, 
\begin{equation}
\label{eq:sum-fractions}
\alpha^2 + \beta^2 = 1; 
\end{equation}
$E_a$ is the magnitude of the electric field of the incident plane wave for $z<0$,
\begin{equation}
\label{eq:E+vec}
\mathbf{E}_a = E_{ax}\, \hat{\mathbf{x}} + 
E_{ay}\, \hat{\mathbf{y}} ;
\end{equation}

Similarly, on the right hand side, $z>0$, we have that 
\begin{equation}
\label{eq:campos-n2}
\mathbf{E}'(z) = 
\left( E'_{bx} \hat{\mathbf{x}} + 
E'_{by} \hat{\mathbf{y}} \right) e^{ik'z} + 
\left( E'_{ax} \hat{\mathbf{x}} + E'_{ay} \hat{\mathbf{y}} 
\right) e^{-ik'z} 
\end{equation}
where $k'=n'\omega/c$ and 
\begin{equation}
\label{eq:alpha'-beta'}
E'_{ax} = \alpha'\, E'_a 
\quad \mbox{and} \quad 
E'_{ay} = \beta'\, E'_a \, e^{i\phi_a'},
\end{equation}
where $\phi'_a$ is the phase difference and $E'_a$ is the magnitude of   
\begin{equation}
\label{eq:E-vec}
\mathbf{E}'_a = E'_{ax}\, \hat{\mathbf{x}} + 
E'_{ay}\, \hat{\mathbf{y}} , 
\end{equation}
and $\alpha'$ and $\beta'$ satisfy an equation equivalent to (\ref{eq:sum-fractions}).

Equations (\ref{eq:alpha-beta}) and (\ref{eq:alpha'-beta'}) can be written in a matrix form 
\begin{equation}
\label{eq:Exy-Dab-Ea}
\left( \begin{array}{c}
E_{ax} \\ E'_{ax}
\end{array} \right) = D_{\alpha}
\left( \begin{array}{c}
E_a \\ E'_a
\end{array} \right) \quad \mbox{and}\quad
\left( \begin{array}{c}
E_{ay} \\ E'_{ay}
\end{array} \right) = D_{\beta}
\left( \begin{array}{c}
E_a \\ E'_a
\end{array} \right),
\end{equation}
where 
\begin{equation}
\label{eq:Da-Db}
D_{\alpha} = \left( \begin{array}{cc}
\alpha & 0 \\ 0 & \alpha' 
\end{array}
\right) \quad\mbox{and}\quad 
D_{\beta} = \left( \begin{array}{cc}
\beta e^{i\phi_a} & 0 \\ 0 & \beta' e^{i\phi'_a} 
\end{array}
\right),
\end{equation}
that, in correspondence with Eq. (\ref{eq:sum-fractions}) must satisfy the following condition 
\begin{equation}
\label{eq:unitary-fractions}
D_{\alpha}^2 + D_{\beta}^{\dagger}D_{\beta} = 
D_{\alpha}^2 + D_{\beta} D_{\beta}^{\dagger} = 
\openone_2.
\end{equation}

What do next is to find the electric fields of the outgoing waves, 
\begin{eqnarray}
\mathbf{E}_b & = & E_{bx}\, \hat{\mathbf{x}} + 
E_{by}\, \hat{\mathbf{y}}, \\  
\mathbf{E}'_b & = & E'_{bx}\, \hat{\mathbf{x}} + 
E'_{by}\, \hat{\mathbf{y}} ,
\end{eqnarray}
in terms of the incoming ones. 

Imposing the boundary conditions at the interface, $z=0$, we obtain Fresnel equations in a matrix form\cite{vdom2008} 
\begin{equation}
\label{eq:Fresnel-eqs-1}
\left( \begin{array}{c}
E_{bx} \\ E'_{bx} \\ \hline E_{by} \\ E'_{by}
\end{array} \right) = 
\left( \begin{array}{c|c}
S_F & 0_2 \\ \hline 0_2 & S_F 
\end{array} \right)
\left( \begin{array}{c}
E_{ax} \\ E'_{ax} \\ \hline E_{ay} \\ E'_{ay}
\end{array} \right) , 
\end{equation}
where $0_2$ is a $2\times 2$ null matrix and $S_F$ is a $2\times 2$ matrix given by 
\begin{equation}
S_F = \left( \begin{array}{cc} 
-r_F & t'_F \\ t_F & -r'_F
\end{array} \right), 
\end{equation}
where, $r_F$ ($r'_F$) and $t_F$ ($t'_F$) are the reflection and transmission Fresnel coefficients for electric field waves incoming from the left (right), whose explicit form, in terms of the indices of refraction, are\cite{Jackson} 
\begin{eqnarray}
\label{eq:Fresnel}
r_F = \frac{n'-n}{n'+n}, & & r'_F = -r_F, \\
\label{ec:cFresnel2}
t_F = \frac{2n}{n'+n}, & & 
t'_F = \frac{2n'}{n'+n}. 
\end{eqnarray}

Following Ref.~\onlinecite{vdom2008} we introduce renormalized electric field amplitudes  given by 
\begin{equation}
\label{eq:E-normal}
\mathcal{E}_{lm}=\sqrt{n}E_{lm}, \quad  
\mathcal{E}'_{lm}=\sqrt{n'}E'_{lm}\quad (l=a,b;\,\, m=x,y).
\end{equation}
Therefore, Eq. (\ref{eq:Fresnel-eqs-1}) can now be rewritten as 
\begin{equation}
\label{eq:Fresnel-eqs-2}
\left( 
\begin{array}{c}
\mathcal{E}_{bx} \\ \mathcal{E}'_{bx} \\ \hline 
\mathcal{E}_{by} \\ \mathcal{E}'_{by}
\end{array} 
\right) = 
S_{xy} \left( 
\begin{array}{c}
\mathcal{E}_{ax} \\ \mathcal{E}'_{ax} \\ \hline 
\mathcal{E}_{ay} \\ \mathcal{E}'_{ay}
\end{array} 
\right) , 
\end{equation}
where 
\begin{equation}
\label{eq:S-xy} 
S_{xy} = \left( 
\begin{array}{c|c}
S_2 & 0_2 \\ \hline 0_2 & S_2 
\end{array} 
\right),
\end{equation}
with 
\begin{equation}
\label{eq:S-linear-1} 
S_2 = \left( \begin{array}{cc}
\sqrt{n} & 0 \\ 0 & \sqrt{n'}
\end{array}
\right) S_F \left( \begin{array}{cc}
\frac{1}{\sqrt{n}} & 0 \\ 0 & \frac{1}{\sqrt{n'}}
\end{array}
\right) .
\end{equation}
being the $2\times 2$ scattering matrix for normal incidence of a single linearly polarized plane wave (it was named $S$ in Ref.~\onlinecite{vdom2008}). 

The scattering matrix $S_{xy}$ defined in Eq. (\ref{eq:Fresnel-eqs-2}), is a $4\times 4$ unitary and symmetric matrix because $S_2$ is itself a $2\times 2$ unitary and symmetric 
matrix.\cite{vdom2008} Therefore, flux conservation is fulfilled as well as time inversion invariance; $S_{xy}$ is block diagonal since it is written in the basis of linear polarization, where the $x$ and $y$ components, called {\em channels} in the nomenclature of nuclear physics, are decoupled from each other. However, at this stage is not completely clear how to recover the simpler result of a single linearly polarized wave;  we remedy this by a change of basis as shown in the next subsection.

\subsection{Rotated basis}

From Eqs. (\ref{eq:Exy-Dab-Ea}) and (\ref{eq:Fresnel-eqs-2}) we have that 
\begin{equation}
\label{eq:Fresnel-eqs-3}
\left( \begin{array}{c}
\mathcal{E}_{bx} \\ \mathcal{E}'_{bx} \\ \hline 
\mathcal{E}_{by} \\ \mathcal{E}'_{by}
\end{array} \right) = 
\left( \begin{array}{c|c}
S_2D_{\alpha} & 0_2 \\ \hline 0_2 & S_2D_{\beta} 
\end{array} \right)
\left( \begin{array}{c}
\mathcal{E}_a \\ \mathcal{E}'_a \\ \hline 
\mathcal{E}_a \\ \mathcal{E}'_a
\end{array} \right) .
\end{equation}
Now, if a $\pi/4$-rotation is applied to the basis used up to now, by means of\cite{Gopar1996}
\begin{equation}
\label{eq:R0}
R_0 = \frac{1}{\sqrt{2}} \left( \begin{array}{cc}
\openone_2 & \openone_2 \\ -\openone_2 & \openone_2
\end{array}\right),
\end{equation}
equation (\ref{eq:Fresnel-eqs-3}) becomes 
\begin{equation}
\label{eq:Fresnel-eqs-4}
\left( \begin{array}{c}
\mathcal{E}_{b1} \\ \mathcal{E}'_{b1} \\ \hline 
\mathcal{E}_{b2} \\ \mathcal{E}'_{b2}
\end{array} \right) = 
S' \left( \begin{array}{c}
\sqrt{2}\mathcal{E}_a \\ \sqrt{2}\mathcal{E}'_a \\ \hline 
0 \\ 0
\end{array} \right) ,
\end{equation}
where we have defined 
\begin{eqnarray}
\label{eq:ampls-1}
\left( \begin{array}{c}
\mathcal{E}_{b1} \\ \mathcal{E}'_{b1}
\end{array} \right) & = & 
\frac{1}{\sqrt{2}} \left( \begin{array}{c}
\mathcal{E}_{by} \\ \mathcal{E}'_{by} 
\end{array} \right) + 
\frac{1}{\sqrt{2}} \left( \begin{array}{c}
\mathcal{E}_{bx} \\ \mathcal{E}'_{bx} 
\end{array} \right),  \\ 
\label{eq:ampls-2} 
\left( \begin{array}{c}
\mathcal{E}_{b2} \\ \mathcal{E}'_{b2}
\end{array} \right) & = & 
\frac{1}{\sqrt{2}} \left( \begin{array}{c}
\mathcal{E}_{by} \\ \mathcal{E}'_{by} 
\end{array} \right) - 
\frac{1}{\sqrt{2}} \left( \begin{array}{c}
\mathcal{E}_{bx} \\ \mathcal{E}'_{bx} 
\end{array} \right),
\end{eqnarray}
and 
\begin{equation}
\label{eq:Fresnel-eqs-5}
S' = R_0\left( 
\begin{array}{c|c}
S_2D_{\alpha} & 0_2 \\ \hline 0_2 & S_2D_{\beta} 
\end{array} 
\right) R_0^T,
\end{equation}
is a new scattering matrix. It is important to realize that $S'$ does not have the familiar form of a scattering matrix, as can be see by (\ref{eq:Fresnel-eqs-4}).

In a standard scattering matrix the 11 and 22 blocks relate the incoming amplitudes on one side to the outgoing ones in the same side, and the 12 and 21 blocks relate the incoming amplitudes on one side to the outgoing ones on the other side.
Since we want to keep the familiar and very useful interpretation of a scattering matrix we transform (\ref{eq:Fresnel-eqs-4}) by means of the following orthogonal transformation matrix 
\begin{equation}
O = \left( 
\begin{array}{cccc}
1 & 0 & 0 & 0 \\ 
0 & 0 & 1 & 0 \\
0 & 1 & 0 & 0 \\
0 & 0 & 0 & 1
\end{array}
\right), 
\end{equation}
that rearranges the incoming vector in a way that can be more easily interpreted in scattering theory. Therefore, we obtain that 
\begin{equation}
\label{eq:S-matrix-1}
\left( \begin{array}{c}
\mathcal{E}_{b1} \\ \mathcal{E}_{b2} \\ \hline
\mathcal{E}'_{b1} \\ \mathcal{E}'_{b2}
\end{array} \right) = S
\left( \begin{array}{c}
\sqrt{2}\mathcal{E}_a \\ 0 \\ \hline \sqrt{2}\mathcal{E}'_a \\ 0
\end{array} \right) ,
\end{equation}
where  
\begin{equation}
\label{eq:S-matrix-2}
S = O R_0 
\left( \begin{array}{c|c}
S_2D_{\alpha} & 0_2 \\ \hline 0_2 & S_2D_{\beta} 
\end{array} \right) 
R_0^T O^T,
\end{equation}
is a scattering matrix whose dimension is twice the number of channels. This scattering matrix can now be directly identified as: 
\begin{equation}
\label{eq:S-matrix-3}
S = \left( \begin{array}{cc}
r & t' \\ t & r' 
\end{array} \right) ,
\end{equation}
where
$r$ ($r'$) and $t$ ($t'$) are immediately identified as the corresponding $2\times 2$ reflection and transmission matrices, respectively, for incidence from the left (right); they are explicitly given by 
\begin{eqnarray}
\label{eq:r-t-matrices}
r & = &  \frac{n-n'}{n+n'} F , \quad 
t = \frac{2\sqrt{nn'}}{n+n'} F, \\ 
\label{eq:rp-tp-matrices}
r' & = & - \frac{n-n'}{n+n'} F' , \quad 
t' = \frac{2\sqrt{nn'}}{n+n'} F',
\end{eqnarray}
where
\begin{eqnarray}
\label{eq:F-matrix}
F & = & \frac 12\left( \begin{array}{cc} 
\beta e^{i\phi_a} + \alpha &  
\beta e^{i\phi_a} - \alpha \\
\beta e^{i\phi_a} - \alpha & 
\beta e^{i\phi_a} + \alpha 
\end{array} \right) , \\
F' & = & \frac 12\left( \begin{array}{cc} 
\beta' e^{i\phi'_a} + \alpha' &  
\beta' e^{i\phi'_a} - \alpha' \\
\beta' e^{i\phi'_a} - \alpha' & 
\beta' e^{i\phi'_a} + \alpha' 
\end{array} \right).
\end{eqnarray}

Even though $S$ satisfies flux conservation, 
\begin{equation}
\label{eq:S-unitary}
S^{\dagger}S=\openone_4,
\end{equation}
does not possesses a time reversal invariance, 
\begin{equation}
S \neq S^T, 
\end{equation}
where the superscript $T$ denotes transposition. Also, since there is no specular symmetry we have that $r'\neq r$. 

We ask the reader to interpret why $S_{xy}$ is symmetric while $S$ is not. (Suggestion: see the discussion leading to the derivation of Stoke's relations.\cite{Marion,Hecht})

The reflection and transmission coefficients are now given by  
\begin{eqnarray}
R & = & \mbox{tr} \left( r\,r^{\dagger} \right) = \frac{(n-n')^2}{(n+n')^2} 
\label{eq:R-def} \\ 
T & = & \mbox{tr} \left( t\,t^{\dagger} \right) = \frac{4nn'}{(n+n')^2} , 
\label{eq:T-def}
\end{eqnarray}
where we used that $\mbox{tr}(FF^{\dagger})=\mbox{tr}(F'{F'}^{\dagger})=1$, 
where we note that due to flux conservation
\begin{equation}
\label{eq:unit1}
R + T = 1. 
\end{equation}
Given all this, we can write the electric fields, Eqs. (\ref{eq:ampls-1}) and (\ref{eq:ampls-2}), as  
\begin{eqnarray}
\label{eq:E-new-basis-1}
\mathbf{E}(z) & = & 
\frac{\sqrt{2}\mathcal{E}_a}{\sqrt{n}}\, \hat{\mathbf{e}}_1\, e^{ikz} + 
\left( \frac{\mathcal{E}_{b1}}{\sqrt{n}}\, \hat{\mathbf{e}}_1 + 
\frac{\mathcal{E}_{b2}}{\sqrt{n}}\, \hat{\mathbf{e}}_2
\right)\, e^{-ikz} \qquad \\ 
\label{eq:E-new-basis-2}
\mathbf{E}'(z) & = & 
\frac{\sqrt{2}\mathcal{E}'_a}{\sqrt{n'}}\hat{\mathbf{e}}_1e^{-ik'z} + 
\left( \frac{\mathcal{E}'_{b1}}{\sqrt{n'}}\hat{\mathbf{e}}_1 + 
\frac{\mathcal{E}'_{b2}}{\sqrt{n'}}\hat{\mathbf{e}}_2
\right)e^{ik'z},
\end{eqnarray}
where we have introduced a new basis set of vectors 
\begin{equation}
\hat{\mathbf{e}}_1 = \frac{1}{\sqrt{2}} 
( \hat{\mathbf{y}} + \hat{\mathbf{x}} ) 
\quad \mbox{and} \quad 
\hat{\mathbf{e}}_2 = \frac{1}{\sqrt{2}}
( \hat{\mathbf{y}} - \hat{\mathbf{x}} ),
\end{equation}
which plays a role equivalent role to the basis used in Ref.~\onlinecite{Jackson} to discuss circular polarization. 

We left to the reader the analysis necessary to verify that $S$ reduces to the linear polarization case studied in Ref. \onlinecite{vdom2008}, as well as to study the circular polarization case.


\section{Dielectric-conductor interface}
\label{sec:conductor}

We now study a different case. Lets assume that the media on the right side of the interface is a conductor with an electric conductivity $\sigma$, such that its refractive index is complex: $n'\rightarrow n'+i\eta'$, where $n'$ and $\eta'$ are the optical constants;\cite{Reitz} the corresponding wave number is also complex. In the treatment of last section we replace $k'\rightarrow k'+i\kappa'$ where 
\begin{equation}
k' = n'\omega/c \quad \mbox{and} \quad \kappa' = \eta'\omega/c.
\end{equation}

On the dielectric side, from Eqs. (\ref{eq:E-new-basis-1}) and (\ref{eq:E-new-basis-2}) with $\mathcal{E}'_a=0$ we have that 
\begin{eqnarray}
\label{eq:E-new-die}
\mathbf{E}(z) & = & 
\frac{\sqrt{2}\mathcal{E}_a}{\sqrt{n}}\hat{\mathbf{e}}_1e^{ikz} + 
\left( \frac{\mathcal{E}_{b1}}{\sqrt{n}}\hat{\mathbf{e}}_1 + 
\frac{\mathcal{E}_{b2}}{\sqrt{n}} \hat{\mathbf{e}}_2
\right) e^{-ikz} \qquad\quad \\ 
\label{eq:E-new-cond}
\mathbf{E}'(z) & = &  
\left( \frac{\mathcal{E}'_{b1}}{\sqrt{n'+i\eta'}} \hat{\mathbf{e}}_1 + 
\frac{\mathcal{E}'_{b2}}{\sqrt{n'+i\eta'}} \hat{\mathbf{e}}_2
\right) e^{-\kappa'z} e^{ik'z}.
\end{eqnarray}

By definition, the scattering matrix relates the outgoing to the incoming plane wave amplitudes; therefore 
\begin{equation}
\label{eq:S-tilde-1}
\left( \begin{array}{c}
\mathcal{E}_{b1} \\ \mathcal{E}_{b2} 
\end{array} \right) = \widetilde{S}
\left( \begin{array}{c}
\sqrt{2}\mathcal{E}_a \\ 0 
\end{array} \right) ,
\end{equation}
where 
\begin{equation}
\label{eq:S-tilde-2}
\widetilde{S} = \frac{n-n'-i\eta'}{n+n'+i\eta'}\, F.
\end{equation}
Note that now $\widetilde{S}$ describes the reflection back to the same side where  the elliptically polarized planes waves are arriving in a normal direction to the interface; therefore, it is a $2\times 2$ matrix with the following structure
\begin{equation}
\tilde{S} = \left(
\begin{array}{cc}
\tilde{r} & \tilde{t}' \\ 
\tilde{t} & \tilde{r}'
\end{array}
\right),
\end{equation}
where now $\tilde{r}$ ($\tilde{r}'$) is the reflection amplitude when incidence is on the channel $\hat{\mathbf{e}}_1$ ($\hat{\mathbf{e}}_2$), and $\tilde{t}$ ($\tilde{t}'$) is the transmission amplitude from channel $\hat{\mathbf{e}}_1$ ($\hat{\mathbf{e}}_2$) to channel $\hat{\mathbf{e}}_2$ ($\hat{\mathbf{e}}_1$). Here, however, $\hat{\mathbf{e}}_1$ is the only channel where incidence is possible; note also that $\tilde{t}'=\tilde{t}$ such that 
\begin{equation}
\widetilde{S} = \widetilde{S}^T,
\end{equation}
since there is time reversal invariance. Since channels $\hat{\mathbf{e}}_1$ and $\hat{\mathbf{e}}_2$ are equivalent to each other, there is reflection symmetry thing for which $\tilde{r}'=\tilde{r}$. It is very important to realize that $\widetilde{S}$ is not unitary, and that according to Eq. (\ref{eq:S-tilde-2}), can be parametrized as 
\begin{equation}
\label{eq:S-tilde-3}
\widetilde{S} = \sqrt{R}\, e^{i\theta}\, F,
\end{equation}
where $R$ is the reflection coefficient given by 
\begin{equation}
\label{eq:R-tilde1}
R = \frac{(n-n')^2+{\eta'}^2}{(n+n')^2+{\eta'}^2},
\end{equation}
and $\theta$ is a measure of the phase shift between incoming and reflected waves, and  satisfies 
\begin{equation}
\label{eq:theta}
\tan\theta = \frac{-2n\eta'}{n^2-{n'}^2-{\eta'}^2}.
\end{equation}

\begin{figure}
\includegraphics[width=0.65\columnwidth]{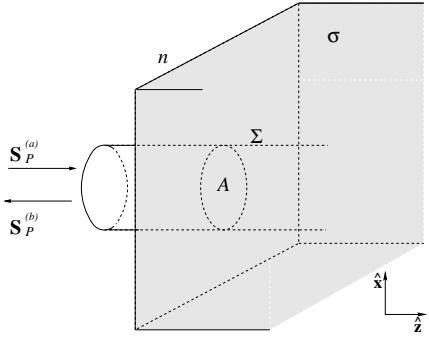} 
\caption{On the right hand side, $z>0$, there is a conductor with conductivity $\sigma$, and for $z<0$ we have a dielectric material. The arrows represent Poynting's vectors corresponding to the electric fields on the dielectric.}
\label{fig:Poynting2}
\end{figure}

Lets see how an energy balance consideration impose restrictions on $\widetilde{S}$. Consider a closed surface $\Sigma$ consisting of a semi-infinite cylinder of cross section $A$, as shown in Fig.~\ref{fig:Poynting2}; in this case, Poynting's vectors are directed in the $z$-direction and the energy flux takes place through $A$ and not through the lateral surface. The energy balance, Eq. (\ref{eq:balance}), is written as
\begin{equation}
\label{eq:balance-conductor}
\Phi_{b1} + \Phi_{b2} - \Phi_a = -W,  
\end{equation}
where, based on (\ref{eq:sprom2}) and (\ref{eq:flux}),\cite{vdom2008} 
\begin{eqnarray}
\label{eq:flux-e1-e2-1}
\Phi_{bj} & = &  
\frac{1}{2\mu_0 c}
\left| \mathcal{E}_{bj} \right|^2 A , 
\quad j=1,\,2, \\ 
\label{eq:flux-e1-e2-2}
\Phi_a & = & \frac{1}{2\mu_0 c}
\left| \sqrt{2} \mathcal{E}_a \right|^2 A,
\end{eqnarray}
From Eq. (\ref{eq:W}), the rate of work made by the field is then given by 
\begin{equation}
W = \frac 12\int_0^{\infty} 
\frac{\sigma}{\sqrt{{n'}^2+{\eta'}^2}} 
\left( \left| \mathcal{E}'_{b1}\right|^2 + 
\left| \mathcal{E}'_{b2}\right|^2 \right) e^{-2\kappa z}Adz ,
\end{equation}
where we use that $\mathbf{J}(z)=\sigma\mathbf{E}'(z)$; integrating we have that 
\begin{equation}
\label{eq:work}
W = \frac{\sigma A}{4\kappa\sqrt{{n'}^2+{\eta'}^2}} 
\left( \left| \mathcal{E}'_{b1} \right|^2 + 
\left| \mathcal{E}'_{b2}\right|^2 \right).
\end{equation}
Using Eqs. (\ref{eq:flux-e1-e2-1}), (\ref{eq:flux-e1-e2-2}) and (\ref{eq:work}), Eq. (\ref{eq:balance-conductor}) gives 
\begin{equation}
\label{eq:Psub1}
( | \mathcal{E}_{b1} |^2 + 
| \mathcal{E}_{b2} |^2 ) -
2 | \mathcal{E}_a |^2 = -
\frac{\sigma\mu_0c}{2\kappa\sqrt{{n'}^2+{\eta'}^2}} 
( | \mathcal{E}'_{b1} |^2 + 
| \mathcal{E}'_{b2} |^2 ),
\end{equation}
which on a matrix form is written as 
\begin{eqnarray}
\label{eq:Psub2}
& & \left( \begin{array}{cc} 
\sqrt{2}\mathcal{E}^*_a & 0 \end{array} \right) 
( \widetilde{S}^{\dagger}\widetilde{S} - \openone_2 ) 
\left( \begin{array}{c} 
\sqrt{2} \mathcal{E}_a \\ 0 \end{array} \right) 
\nonumber \\ & & = -
\frac{\sigma\mu_0c}{2\kappa\sqrt{{n'}^2+{\eta'}^2}} 
\left( \begin{array}{cc} 
{\mathcal{E}'}^*_{b1} & {\mathcal{E}'}^*_{b2} 
\end{array} \right)
\left( \begin{array}{c} 
\mathcal{E}'_{b1} \\ \mathcal{E}'_{b2} 
\end{array} \right).
\end{eqnarray}
Now, using Eq. (\ref{eq:S-matrix-1}) with $\mathcal{E}'_a=0$ and $n'\rightarrow n'+i\eta'$ we can write that 
\begin{equation}
\left( \begin{array}{c} 
\mathcal{E}'_{b1} \\ \mathcal{E}'_{b2} 
\end{array} \right) = t 
\left( \begin{array}{c} 
\sqrt{2}\mathcal{E}_a \\ 0 
\end{array} \right),
\end{equation}
where [see Eqs. (\ref{eq:r-t-matrices})]
\begin{equation}
t = \frac{2\sqrt{n(n'+i\eta')}}{n+n'+i\eta'} \, F.
\end{equation}
Therefore, Eq. (\ref{eq:Psub2}) is equivalent to
\begin{equation}
\label{eq:S-tilde-unitarity}
\widetilde{S}^{\dagger}\widetilde{S} + t_p^{\dagger}t_p = \openone_2,
\end{equation}
where 
\begin{equation}
t_p = \sqrt{\frac{\sigma\mu_0c}{2\kappa\sqrt{{n'}^2+{\eta'}^2}}}\, t ,
\end{equation}
which can also be written as
\begin{equation}
t_p = \sqrt{T_p}\, e^{i\phi} \, F, 
\end{equation}
where now 
\begin{equation}
\label{eq:T_p}
T_p = \frac{4nn'}{(n+n')^2+{\eta'}^2},
\end{equation}
and $\phi$ is given by
\begin{equation}
\tan\phi = -\frac{\eta'}{n+n'}.
\end{equation}

Calculating the trace of Eq.~(\ref{eq:S-tilde-unitarity}) we obtain an expression that in short is equivalent to energy flux conservation 
\begin{equation}
R + T_p = 1,
\end{equation}
where $R$, the reflection coefficient is the fractional amount of energy that is turned back into the dielectric, while $T_p$ tell us the amount of energy dissipated or  ``lost'' in the conductor to non propagating modes or, what we call {\em parasitic channels}.\cite{Lewenkopf1992,vdom2008}

Hence, Eq.~(\ref{eq:S-tilde-unitarity}) can be seen as the unitarity condition for a $4\times 4$ scattering matrix $S$ that satisfies flux conservation, and is given by 
\begin{equation}
\label{eq:Sparasitic}
S = \left(
\begin{array}{cc}
\tilde{S} & t_p \\ t_p & S_{pp} 
\end{array}
\right),
\end{equation}
where the unitarity of $S$ implies that  
\begin{equation}
S_{pp} = -\frac{1}{t_p^{\dagger}}\tilde{S}^{\dagger}t_p =
 -\sqrt{1-T_p} e^{i(2\phi-\theta)} F .
\end{equation}
The lack of unitarity of  $\widetilde{S}$ is then given by following expression 
\begin{equation}
P = I_2 - \tilde{S}^{\dagger}\tilde{S} = t_p^{\dagger}t_p, 
\end{equation}
whose trace quantifies ohmic losses through an {\em absorption strength} parameter 
\begin{equation}
\gamma = \mbox{tr}\, P = T_p,
\end{equation}
where it must be emphasized that an ohmic loss takes place through parasitic channels in the language employed in scattering theory. 

Therefore, the $S$-matrix given by Eq. (\ref{eq:Sparasitic}) describes how the energy losses in the conductor can be interpreted as energy lost to parasitic channels characterized by the $t_p$ matrix, where the coupling between the two media is described by the material constants of our model; in particular, in this work, for the elliptic polarization case. Even though we have two parasitic channels the total absorption is measured by a single parameter $\gamma=T_p$, that gives the same result previously obtained in the linear polarization case, something expected since each polarization mode contributes just a fraction of the total energy flux.


\end{document}